\documentclass[aps,floats,epsf, twocolumn]{revtex4}
 \usepackage{graphics}

\begin{document}

\title{Spectrum of the Dirac Hamiltonian with the mass-hedgehog in arbitrary dimension}

\author{Igor F. Herbut and Chi-Ken Lu}

\affiliation{Department of Physics, Simon Fraser University, Burnaby, British Columbia, Canada V5A 1S6}

\begin{abstract}
It is shown that the square of the Dirac Hamiltonian with the isotropic mass-hedgehog potential in $d$ dimensions is the number operator of fictitious bosons and fermions over $d$ quantum states. This result allows one to obtain the complete spectrum and degeneracies of the Dirac Hamiltonian with the hedgehog mass configuration in any dimension. The result pertains to low-energy states in the core of a general superconducting or insulating vortex in graphene in two dimensions, and in the superconducting vortex at the topological - trivial insulator interface in three dimensions, for example. The spectrum in $d=2$ is also understood in terms of the underlying accidental $SU(2)$ symmetry and the supersymmetry of the Hamiltonian.
\end{abstract}
\maketitle

\vspace{10pt}

\section {Introduction}

The spectra of Dirac particle in topologically non-trivial backgrounds have been attracting attention ever since the pioneering work of Jackiw and Rebby on fractionalization of electric charge.\cite{jackiw} With the recent rise of graphene and topological insulators this problem has only gained in importance. The most interesting feature of the spectrum probably is the appearance of states with precisely zero-energy, which are topologically protected and have several exotic manifestations.\cite{kopnin, volovik, hou, herbut1, pi, fu, wilczek,ghaemi,khyamovich, lehur,  herbut2, teo}  The best understood example is provided by the two-dimensional Dirac equation in the vortex background, which in its various representations is applicable to both graphene\cite{hou, herbut1, pi, wilczek, ghaemi, khyamovich, lehur, herbut2} and the surface of a topological insulator\cite{fu,teo}. Very recently, it was shown\cite{teo} that the unique zero-energy state also exists in the three-dimensional version of this Hamiltonian, in which the mass-vortex is replaced by its three-dimensional topological equivalent, the hedgehog. This Hamiltonian describes, for example, a superconductor coexisting with the interface between a topological and a trivial insulator, and its three-dimensional zero-modes were related to non-Abelian statistics for such topological defects.\cite{teo}

Unlike the zero-modes, the rest of the energy spectrum of topologically non-trivial Dirac Hamiltonians has seen  relatively little study \cite{babak, khyamovich}. In this paper we consider the spectrum of the Dirac Hamiltonian in the simplest topologically non-trivial background of a hedgehog in general number of $d$ dimensions, in which the boundary of the configurational space wraps the $S_{d-1}$ sphere of the order parameter (i. e. masses of Dirac fermion) space fully once. We show that the simplest Dirac Hamiltonian of this kind is in fact a square-root of the particle number operator for a collection of fictitious  bosons and fermions distributed over $d$ quantum states:
  \begin{equation}
  (2 \hat{N}) ^{1/2} = \vec{\alpha} \cdot \hat{\vec{p}} + \vec{\beta} \cdot \hat{\vec{r}}.
  \end{equation}
  Here $\hat{N}= \sum_{i=1}^d (\hat{n}_i ^b + \hat{n}_i ^f )$ is the sum of the standard bosonic and fermionic number operators, $\hat{\vec{p}}$ and $\hat{\vec{r}}$ are the momentum and position operators,
  $[\hat{r}_i, \hat{p} _j ] = i \delta_{ij} $, and $\alpha_i$ and $\beta_i$, $i=1,...d$ are the usual Hermitian Dirac matrices that satisfy the $2d$-dimensional Clifford algebra,  $ \{ \alpha_i , \alpha_j \} = \{ \beta_i , \beta_j \}= 2 \delta_{ij}$,
  $\{ \alpha_i , \beta_j \}=0$. This surprisingly simple formula, which may be understood as a generalization of the celebrated Dirac connection between the bosonic number operator and the quantum harmonic oscillator, enables one to determine the spectrum and the degeneracies of the Dirac Hamiltonian on the right hand side by purely combinatorial or algebraic means. In particular, it immediately follows that there is a non-degenerate zero-energy state in {\it any} dimension, which persists under smooth deformations of the mass-hedgehog configuration in Eq. (1). The above Hamiltonian is particularly relevant in two dimensions, since there it represents the leading term in the expansion of any realistic Dirac vortex Hamiltonian near its center, and the spectrum that will be obtained here may be understood as the first approximation to the one of the real system at energies close to zero. We therefore also provide an alternative understanding of the spectrum in terms of symmetries of the Hamiltonian, and discuss some consequences of these considerations. In particular, we show that the mathematical origin of the observed degeneracies are {\it two} separate accidental symmetries of the Hamiltonian: one closely related to the accidental symmetry of the harmonic oscillator, and the second that may be understood as a form of supersymmetry.\cite{witten}

     The paper is organized as follows. In the next section we first motivate and then define the hedgehog Dirac Hamiltonian. We then proceed to compute its spectrum in arbitrary dimension by introducing suitable bosonic and {\it fermionic}  creation and annihilation operators.  The anisotropic version of the Hamiltonian is solved  in sec. III. In sec. IV we discern a hidden $SU(2)$ symmetry of the Hamiltonian, which explains a part of the observed degeneracies. In sec. V we present a set of operators which close the Clifford algebra, and which are responsible for the additional degeneracy of the spectrum. We close with the remark on applicability of our results to more realistic Hamiltonians that would represent vortices in systems with Dirac fermions, other comments, and the short summary.

\section{Bose and fermi operators and the spectrum of the hedgehog}

  To motivate what follows let us begin with the familiar Dirac Hamiltonian in a mass-vortex background in two dimensions \cite{kopnin, volovik, hou, herbut1, pi, fu, wilczek,ghaemi,khyamovich, lehur,  herbut2}:
 \begin{equation}
 \hat{H}_0 = c (\alpha_1 \hat{p}_1  + \alpha_2 \hat{p}_2)   +   |\Delta(r)| ( \beta_1 \cos \theta   + \beta_2 \sin \theta   ) ,
 \end{equation}
 where $\alpha_i$ and $\beta_i$, $i=1,2$, are four-dimensional, unitary, Hermitian, anticommuting matrices, and $(r,\theta)$ are the usual polar coordinates. Finiteness  of the energy of the vortex configuration, as usual, requires that the amplitude $|\Delta(r)| \rightarrow 0$ when $r\rightarrow 0$. In spirit of the harmonic approximation, expanding  $|\Delta(r)|$ near the origin and retaining only the leading term suggests an introduction of  the {\it Dirac hedgehog Hamiltonian} as
  \begin{equation}
 \hat{H} = c \vec{\alpha} \cdot \hat{\vec{p}} +  m \vec{\beta} \cdot \hat{\vec{r}} ,
 \end{equation}
 in {\it general number} of spatial dimensions $d$. The last term represents a topologically non-trivial mass-hedgehog configuration in $d$ dimensions, and in particular in $d=2$, evidently $\hat{H}_0 = \hat{H} + O(r^2)$. Here $c$ is a velocity, and $m$ is a parameter with the dimension of energy/length. We work in units in which $\hbar=1$, as usual, and, without loss of generality, will assume $m>0$. Since in general the $2d$  Hermitian matrices $\alpha_i$ and $\beta_i$ form the Clifford algebra $C(2d,0)$, their smallest irreducible representation will be $2^d$-dimensional.

 Squaring the Hamiltonian $\hat{H}$ yields
\begin{equation}
\hat{H}^2 = c^2 \hat{\vec{p}} ^2  + m^2 \hat{\vec{r}} ^2  + i m c \vec{\beta}\cdot  \vec{\alpha}.
\end{equation}
Following Dirac, the bosonic operators may be defined as usual:
\begin{equation}
\hat{b}_i = (m/2c)^{1/2} \hat{r_i} + i (c/2m)^{1/2} \hat{p}_i,
\end{equation}
so that $[\hat{b}_i, \hat{b}_j ^\dagger ] = \delta_{ij}$, and the square of the Hamiltonian may be rewritten as
\begin{equation}
\hat{H}^2 = mc \sum_{i=1} ^d  ( 2 \hat{b}_i ^\dagger \hat{b}_i + 1  + i \beta_i \alpha_i).
\end{equation}
To discern fermions in the problem one may recall that the creation and the annihilation operators for Majorana fermions \cite{wilczek} satisfy precisely the same Clifford algebra as the Dirac matrices. Therefore, if we define  a set of new operators as
\begin{equation}
\hat{a}_i = (\beta_i + i \alpha_i)/2,
\end{equation}
the anticommutation rules for the Dirac matrices imply that these linear combinations act as fermionic annihilation and creation operators, since evidently $\{ \hat{a}_i, \hat{a}_j \}= 0$, and  $\{ \hat{a}_i, \hat{a}_j ^\dagger \}= \delta_{ij}$. Furthermore, the product of the Dirac matrices appearing in $\hat{H}^2$ can be simply written in terms of the same fermions as
\begin{equation}
i\beta_i \alpha_i = 2 \hat{a}_i ^\dagger \hat{a}_i - 1 .
\end{equation}
Inserting this into Eq. (6) the oscillator's zero-point energy is canceled and
\begin{equation}
\hat{H} ^2 = 2 m c \sum_{i=1} ^ d   ( \hat{b}_i ^\dagger \hat{b}_i + \hat{a}_i ^\dagger \hat{a}_i),
\end{equation}
as announced in the introduction for $m=c=1$.

 The spectrum of the hedgehog-Dirac Hamiltonian is therefore
\begin{equation}
E_N ^\pm = \pm \sqrt{2 N m c }
\end{equation}
where $N=0,1,2 ...$ is an integer quantum number.  This agrees with \cite{babak} where the spectrum was obtained by solving the differential equation in coordinate representation. There is a non-degenerate state with precisely zero-energy, when the number of both bosons and fermions vanishes. Otherwise, the degeneracy of the eigenvalue $E_N ^\pm$ is
\begin{equation}
D(d,N) = \frac{1}{2} \sum_{n=0} ^d \frac{d!}{(d-n)! n!} \frac{(d+N-n-1)!}{(d-1)! (N-n)! }.
\end{equation}
The two factors under the sum represent the number of ways in which $n$ fermions and $N-n$ bosons can be distributed over $d$ quantum states, respectively. The factor of $1/2$ in front originates in the ``chiral" symmetry between the positive and negative energy eigenstates, implied by the existence of the Hermitian operator
\begin{equation}
\gamma=\prod_{i=1}^d i \alpha_i \beta_i,
\end{equation}
which {\it anticommutes} with the Hamiltonian $\hat{H}$.

In the physical dimensions $d=1,2,3$ for $N\neq 0$ we this way find:
\begin{equation}
D(1,N)=1,
\end{equation}
\begin{equation}
D(2,N)=2N,
\end{equation}
\begin{equation}
D(3,N)=2N^2 +1,
\end{equation}
with the degeneracy in $d$ dimensions being a polynomial function of $N$ of the degree $d-1$. In d=1 all the levels are non-degenerate. For a  general dimension one can also write,
\begin{equation}
D(d,N) = \frac{ (N+d-1)!}{2 N! (d-1)!}F ( -d,-N,1-d-N,-1)
\end{equation}
where $F$ is the hypergeometric function $_2 F_1$.

 It also transpires that the eigenstates including the zero-mode can be chosen to be real. The Clifford algebra $C(n,m)$ for $n+m$-odd has a real representation only if $n-m=1$,  \cite{okubo} with the dimension $2^d$. The Hermitian matrices $\{\alpha_i, \beta_i, \gamma \}$, $i=1,... d$  can thus always be chosen so that $d+1$ of them are purely real, and the remaining $d$ purely imaginary.\cite{lu} Let us therefore chose $\alpha_i$ to be imaginary and $\beta_i$ to be real. The time-reversal operator that commutes with the real Hamiltonian in this representation is simply that of complex conjugation, and therefore {\it all} the eigenstates can be chosen to be real.

\section{Spectrum of the anisotropic hedgehog}

 One can determine the spectrum of a more general Hamiltonian, used for example in ref. \cite{teo}
 \begin{equation}
 \hat{H}= c \vec{\alpha} \cdot \hat{\vec{p} } +  \vec{\beta} \cdot \vec{n}( \hat{\vec{r} } ),
 \end{equation}
 with $n_i = M_{ij} \hat{r} _j$, using a similar technique. We assumed here the velocity to be isotropic, but it is easy to see that even if that would not be the case a simple rescaling would map the problem onto the Hamiltonian in the last equation. Squaring the Hamiltonian now yields
 \begin{equation}
\hat{H}^2 = \sum_{i=1}^d   (c^2 \hat{p}_i \hat{p}_i +  \hat{r}_i (M^\top M)_{ij} \hat{r}_j + i c \beta_i M_{ij} \alpha_j ) .
\end{equation}
A general matrix $M$ can be written in the diadic form \cite{herb-jur-roy} as
\begin{equation}
M= \sum_{i=1}^d m_i (\nu_i  \otimes \mu_i ^\top),
\end{equation}
with $\mu_i ^\top = (\mu_{1i}, \mu_{2i},...\mu_{di})$, $i=1,...d$ as the orthonormal eigenbasis and $\{m_i ^2 \}$ the spectrum of the symmetric matrix $M^\top M$:
\begin{equation}
M^\top M = \sum_{i=1}^d m_i ^2  (\mu_i  \otimes \mu_i ^\top).
\end{equation}
We may therefore still define the bosonic operators as
\begin{equation}
\hat{b}_i = (|m_i|/2c )^{1/2} \hat{\tilde{r}}_i + i (c/2|m_i|)^{1/2} \hat{\tilde{p}}_i,
\end{equation}
and, for $m_i >0$, the fermionic as
\begin{equation}
\hat{a}_i = (\tilde{\beta}_i + i \tilde{\alpha}_i )/2,
\end{equation}
where $\hat{\tilde{r}}_i = \sum_k \mu_{ki} \hat{r}_k$, $\hat{\tilde{p}}_i = \sum_k \mu_{ki} \hat{p}_k$, $\tilde{\alpha}_i = \sum_k \mu_{ki} \alpha_k$, and
$\tilde{\beta}_i = \sum_k \nu_{ki} \beta_k$. If $m_i <0$, $\tilde{\alpha}$ and $\tilde{\beta}$ matrices need to be exchanged in the definition of the fermionic operator. In terms of these ``particles" one finds that
\begin{equation}
\hat{H}^2 = 2 c \sum_{i=1} ^ d  |m_i|  ( \hat{b}_i ^\dagger \hat{b}_i + \hat{a}_i ^\dagger \hat{a}_i).
\end{equation}
Evidently, $\hat{H}$ still has a non-degenerate zero-mode, but the rest of the spectrum becomes modified into
\begin{equation}
E_N ^\pm = \pm (2 c \sum_{k=1} ^d N_k |m_k| )^{1/2},
\end{equation}
where $N_k$ is an integer, including zero.  The degeneracy for a generic set of incommensurate $\{m_k \}$ of these eigenvalues is now at most $2^{d-1}$.

In terms of the particle operators the Hamiltonian itself takes the form
\begin{equation}
\hat{H}= (2 c) ^{1/2} \sum_{i=1}  ^d  |m_i|^{1/2} (\hat{a}_i ^\dagger \hat{b}_i + \hat{b}_i ^\dagger \hat{a}_i),
\end{equation}
from which it is clear that it commutes with the total number of bosons and fermions {\it in each state}. To obtain the eigenvectors therefore one needs to diagonalize a matrix of typical size $2^d \times 2^d$, or $2 D(d,N) \times 2 D(d,N)$, in the isotropic case $m_i =m$.

\section{Accidental SU(2) symmetry in $d=2$}

 Let us now focus on the physically most obviously relevant two-dimensional case, $d=2$, and understand the obtained degeneracies in the isotropic case in a different way.  First, the complete vortex Hamiltonian $\hat{H}_0$ in Eq. (2) is invariant under rotations of the coordinate frame generated by $2 \hat{J}_2 $, where
  \begin{equation}
  \hat{J}_2 = \hat{L}_2 + \hat{S}_2
  \end{equation}
  with $\hat{L}_2 = \hat{\vec{r} } \times \hat{\vec{p}}/2 $ and $ \hat{S}_2 = i (\alpha_2 \alpha_1 + \beta_2 \beta_1)/4$.  In terms of the particle operators,
  \begin{equation}
  \hat{X}_2 = i (\hat{c}_2 ^\dagger \hat{c}_1 - \hat{c}_1 ^\dagger \hat{c}_2)/2,
  \end{equation}
  where $c=b$ (boson) for $X=L$, and $c=a$ (fermion) for $X=S$. Of course, the hedgehog Hamiltonian $\hat{H}$ is  rotationally invariant as well. However, since $\hat{H}$ also commutes with the total particle number in each state it also commutes with their difference:
  $[\hat{H},\hat{J}_3] = 0$, with $\hat{J}_3 = \hat{L}_3 + \hat{S}_3$, and
  \begin{equation}
  \hat{X}_3 = (\hat{c}_1 ^\dagger \hat{c}_1 - \hat{c}_2 ^\dagger \hat{c}_2)/2.
  \end{equation}
  This is an {\it accidental} degeneracy, which originates in the linear dependence of the hedgehog Hamiltonian on coordinate and momentum.  By Jacobi's identity $\hat{H}$ then also commutes with the commutator of the (exact) generator $\hat{J}_2$ and the (accidental) generator $\hat{J}_3$, which happens to be non-trivial: $\hat{J}_1 = \hat{L}_1 + \hat{S}_1$, where
  \begin{equation}
  \hat{X}_1 = (\hat{c}_1 ^\dagger \hat{c}_2 + \hat{c}_2 ^\dagger \hat{c}_1)/2.
  \end{equation}
  Since, by our definition,
\begin{equation}
\hat{X}_i= \sum_{\alpha, \beta=1}^2 \hat{c}^{\dag}_{\alpha}     \sigma^i _{\alpha\beta}    \hat{c}_{\beta}/2,
\end{equation}
where $\sigma^i$ are Pauli matrices, it is easy to see that these operators close the algebra $SU(2) \times SU(2)$:
  \begin{equation}
  [\hat{X}_i, \hat{Y}_j ] = i  \epsilon _{ijk} \hat{X}_k   \delta_{XY},
  \end{equation}
 where $X,Y = L, S$.

We can therefore diagonalize the Hamiltonian $\hat{H}$, the Casimir operator $\hat{\vec{J}^2 } $, and one $\hat{J}_k$ together, as usual. For convenience we will choose here the latter to be the accidental generator $\hat{J}_3$. (One can of course also choose the customary exact rotational generator $\hat{J}_2$, which becomes analogous to our choice after a certain redefinition of the particle operators \cite{messiah}.) The source of degeneracy will now turn out to be twofold: first, there is the ``usual" accidental degeneracy of the harmonic oscillator due to different allowed quantum numbers $m$ (the eigenvalues of $\hat{J}_3$) for a given value of $j$ (where the eigenvalues of $\hat{\vec{J}^2}  $ are $j(j+1)$). This is a consequence of the accidental $SU(2)$ symmetry.  Second, there is an extra degeneracy due to fermions, which can be understood as different allowed values of $j$ at given energy. We will see in the next section that this degeneracy is implied by a hidden supersymmetry of the hedgehog Dirac Hamiltonian. For now, let us simply ask which values of the quantum number $j$ can be present among the eigenstates with the energy $E_N ^+ = \sqrt{2 N mc }$.  It is useful again to consider $\hat{H}^2$ rather than $\hat{H}$, because of its larger symmetry:
\begin{equation}
[\hat{H}^2, \hat{\vec{L}} ]=[\hat{H}^2, \hat{\vec{S}}] =0.
\end{equation}
We can therefore label the eigenstates of $\hat{H}^2$ with the quantum number $N$ with {\it four} additional quantum numbers: $\{ l, m_l, s, m_s\}$, in standard angular momentum notation. In this basis $\hat{H}$ itself is represented by a purely off-diagonal matrix.  We first note that from the definition of $\hat{L}_i$ it readily follows that
\begin{equation}
\hat{\vec{L}^2 } = (\hat{n}_{1} ^b + \hat{n}_{2}^b)(\hat{n}_{1} ^b + \hat{n}_{2} ^b  +2)/4,
\end{equation}
with $\hat{n}_{k} ^b = \hat{b}_k ^\dagger \hat{b}_k $, and therefore we can recognize the quantum number $l$ as being
\begin{equation}
l= N^ b / 2,
\end{equation}
where $N^b = N_{1} ^b  + N_{2} ^b $ is the total number of bosons in a given state.  Note that this reproduces the well-known result \cite{messiah} that the degeneracy of the $E= (2N^b +1)$ level of  $d=2$ harmonic oscillator is $2l+1= N^b+1$. For the fermions, on the other hand,
\begin{equation}
\hat{\vec{S}^ 2 }  = 3( \frac{ \hat{n}_{1} ^f - \hat{n} _{2} ^f }{2} )  ^2.
\end{equation}
We can therefore also write that $s(s+1)= 3 m_s ^2$,
and as there are only two possible values of $m_s ^2$ in the problem, namely, $0$ and $1/4$, in fact $s=|m_s|$. Since this number is finite only when the total number of fermions $N^f = N_{1} ^f + N_{2}^f =1$, and zero when  $N^f =0$  or $N^f =2$, we can finally write
\begin{equation}
s = N^f (2-N^f)/2.
\end{equation}

Since for $N\geq 2$, $l\geq s$, and $s=0, 1/2$  one can write all the allowed values of the the total quantum number $j$ as
  \begin{equation}
  j= l\pm s  = ( N-N^f \pm N^f (2-N^f) )/2 .
  \end{equation}
In spite of the three possible values of the fermion number $N^f$ and two values of the sign in the last equation, for the total number of particles $N\geq 2$ there are in fact only {\it two} values of $j$ allowed:
 \begin{equation}
 j_> = N/2 ,
 \end{equation}
 \begin{equation}
 j_< = (N/2)  -1.
 \end{equation}
 The degeneracy of the energy level $E_N ^\pm = \pm \sqrt{2N}$ is therefore
 \begin{equation}
 D(2,N) = (2 j_> + 1) + (2 j_< + 1 )= 2N.
 \end{equation}
  For $N=1$, on the other hand, it is either that $l=0$, $s=1/2$,  or $l=1/2$, $s=0$. In either case $j=1/2$, so $D(2,1) = 2$. This way the Eq. (14) becomes correct for all $N\neq 0$. The obtained spectrum and the degeneracies are represented in Fig. 1.

\begin{figure}[t]
{\centering\resizebox*{70mm}{!}{\includegraphics{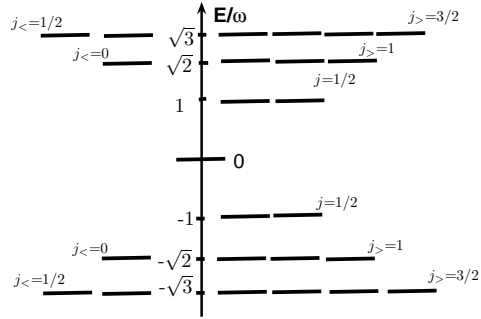}}
\par} \caption[] {The bound state spectrum of  hedgehog Dirac Hamiltonian in $d=2$. The energy is given by $E_{N}/\omega =\pm \sqrt{N}$  where  $\omega= \sqrt{2mc} $. For the energy levels with $N>1$, there are
two degenerate multiplets with the quantum number $j=j_>$ and $j=j_<$ given by Eqs. (38) and (39).}
\end{figure}

\section{Hidden supersymmetry }

The appearance of two values of the quantum number $j$ at the same energy is not explained by the accidental $SU(2)$ symmetry from the last section. Although accidental, its role in the present problem is analogous to the role played by the exact rotational symmetry in the spectrum of the hydrogen atom. The reader may recall that the appearance of different values of the angular quantum number in that case was a consequence of the existence of another conserved quantity in the problem, namely, of the Runge-Lenz-Pauli vector \cite{jauch}. We now show that a similar vector operator exists for the hedgehog Dirac Hamiltonian, and that it may be used to understand the degeneracies, and even derive the energy eigenvalues by purely algebraic means.

Let us now  for simplicity set $2mc =1$, and define three additional Hermitian operators $\hat{A}_i$, $i=1,2,3$, which also obey  $\hat{A}_i ^2=\hat{H}^2$ (without the summation convention over the repeated indices):
\begin{equation}
\hat{A}_i=\sum_{\alpha, \beta=1}^2 \sigma^i _{\alpha\beta} (\hat{a}^{\dag}_{\alpha}\hat{b}_{\beta}
    +\hat{a}_{\beta}\hat{b}^{\dag}_{\alpha}).
\end{equation}
The following algebra of commutation and anticommutation relations is then readily found:
\begin{equation}
\{\hat{A}_i,\hat{A}_j\} = 2\hat{H} ^2\delta_{ij},
\end{equation}
\begin{equation}
[\hat{A}_i,\hat{J}_j] = i\epsilon_{ijk}\hat{A}_k,
\end{equation}
\begin{equation}
\{\hat{H},\hat{A}_i \} = 4\hat{J}_i.
\end{equation}
It is relatively easy to see that these relations then imply that $[\hat{H},\hat{J}_i]=0$, and $[\hat{J}_i, \hat{J}_j] = i \epsilon_{ijk} \hat{J}_k$. The Eq. (42) says that within each eigenspace of $\hat{H}^2$ the operators $\hat{A}_i$ close a Clifford algebra, and Eq. (43) says that $\hat{\vec{A}}$ is indeed a vector under the accidental $SU(2)$ symmetry from the last section. Note also that the chirality operator $\gamma$ satisfies,
\begin{equation}
\{ \gamma, \hat{A}_i \} = [\gamma, \hat{J }_i ]=0.
\end{equation}

We may then use the above algebra to demonstrate that in each energy subspace there could be at most two different values of the quantum number $j$, and that when both of the values coexist they will differ by unity. From Eq. (43) it follows that either
\begin{equation}
\hat{A}_+ | E^2 , j, m\rangle \propto |E^2, j', m+1\rangle,
\end{equation}
or the left hand side is zero, where $\hat{A}_+ = \hat{A}_1 + i \hat{A}_2$, and the first label of the state is the value of $\hat{H}^2$, not of $\hat{H}$. Therefore by choosing $m=j$ we may immediately conclude that $j' \geq j+1$. Since from Eq. (43) we also have $[\hat{J}_+, \hat{A}_+]=0$, it follows that, if not zero, then
\begin{equation}
\hat{ A} _+ | E^2 , j, j\rangle \propto |E^2, j+1, j+1\rangle,
\end{equation}
i. e. in this case it must be that in fact $j' =j+1$.  Finally, since from Eq. (42) it is also true that $\hat{A}_+ ^2 = 0$, it follows that in particular,
\begin{equation}
 \hat{A}_+ ^2 | E^2 , j, j\rangle =0.
\end{equation}
So in each eigenspace of $\hat{H}^2$ there is either only one, or two allowed values of the quantum number $j$. Recalling that $\{ \gamma, \hat{H} \}=0$, from Eq. (45) it readily follows the same is true not only for $\hat{H}^2$, but for the Hamiltonian $\hat{H}$ itself.

 To derive the spectrum from the above algebra, consider the state $|E,j_>, j_> \rangle$, where the first label is now the eigenvalue of the Hamiltonian $\hat{H}$. Using Eq. (44) it is easy to see that
 \begin{equation}
 \langle E,j_>, j_> | \hat{A}_3 |E,j_>, j_> \rangle = \frac{2j_>}{E}.
 \end{equation}
 On the other hand since $[\hat{A}_3, \hat{J}_3] =0$, the operator $\hat{A}_3$ has finite matrix elements  only between the states with the same eigenvalue of $\hat{J}_3$. If that eigenvalue is $j_>$ there are only two such states, and therefore in principle
 \begin{equation}
 \hat{A}_3 |E,j_>, j_> \rangle = \frac{2j_>}{E} |E,j_>, j_> \rangle  + x |-E,j_>, j_> \rangle,
 \end{equation}
 where $x$ is an unknown coefficient. On the other hand, Eq. (45) implies that
 \begin{equation}
 \langle -E, j_>, j_> | \hat{A}_3 |E,j_>, j_> \rangle = - \langle E,j_>, j_> | \hat{A}_3 |-E,j_>, j_> \rangle
 \end{equation}
 i. e. that the coefficient $x$ is purely imaginary. Since on the other hand, the Hamiltonian $\hat{H}$ written in the form in Eq. (25) is manifestly real, the coefficient $x$ must be real as well, and so we conclude that $x=0$.

Finally, since $\hat{A}_3 ^2 = \hat{H}^2$, we may write an equation for the allowed energy eigenvalues as
 \begin{equation}
 E^2 = (\frac{2j_>}{E})^2,
 \end{equation}
 and recover the energy spectrum as $E= \pm \sqrt{ 2 j_>}$ (in our current units $2mc=1$), with the degeneracy $2 j_> +1 + 2 (j_> -1)+1$ for $j_> \geq 1$, and $2 j_> + 1$ for $j_> <1$, where $j_>$ assumes all the allowed values, $j_> =0, 1/2, 1, 3/2, ...$, as usual.

 \section{Remarks}

 If the mass order parameter is isotropic but it saturates, for example,  $n_i (\vec{r}) = m r_i$ for $r<L$,
and $n_i (\vec{r}) = mL$ for $r>L$ in Eq. (17), one can think of the spectrum derived here as being approximately correct if the size of the defect's core $L \gg \sqrt{c/m}$. This is because the low-energy eigenstates, with the energies $\sim \sqrt{mc}$ much below the continuum threshold at $\sim mL$, are Gaussian functions exponentially localized within the length scale of order $\sqrt{c/m}$, and as such will be little affected by the modification of the Hamiltonian far from the center of the vortex. We may thus expect to find $D(d,N)$ levels for small $N$ close together in energy. The zero-energy state, however, being non-degenerate, will still remain at zero in order to preserve the exact reflection symmetry between the positive and negative energy eigenstates of the Hamiltonian.

 Since in $d=2$ the Hamiltonian studied here can be understood as the leading term of a general Dirac vortex Hamiltonian near the origin, one may wonder how much of the degeneracy we identified will remain once the higher order terms in the expansion are included. Non-linear terms in coordinate will of course break the accidental degeneracy, leaving only the $U(1)$ rotational sub-symmetry. Since, as already discussed, by choosing imaginary $\vec{\alpha}$ and real $\vec{\beta}$ the vortex Hamiltonian becomes real, the time-reversal operator is trivial, and by itself does not imply any residual degeneracy. We therefore expect the higher order terms to resolve the degeneracy completely. In the limit of large core where the linear approximation is adequate, the low-energy states should appear in  bundles of $2N$ states grouped together. For a vortex in the superconducting \cite{wilczek, ghaemi, khyamovich, lehur,herbut2}, N\'{e}el \cite{herbut1} or Kekule \cite{hou, roy}, order parameter in graphene, the presence of spin degree of freedom of course doubles these near-degeneracies. In this case even the rotational symmetry is only approximate \cite{herbut3}, and the terms quadratic in momentum reduce it to $C_3$.

 It may also be interesting to note the transformation properties of the generators of the $SU(2)$ symmetry under time reversal. Since under time reversal the bosonic operators stay invariant, it is easiest to choose the representation of the Dirac matrices in which the same is true for the fermionic operators: this is again all $\beta_i$ real and all $\alpha_i$ imaginary. Time reversal in this representation is just the operation of complex conjugation, and whereas the generator $J_2$ of real-space rotation is as usual odd, the accidental generators $J_1$ and $J_3$ are {\it even}. This transformation property is different from that of the true angular momentum, but still perfectly consistent with the enclosed $SU(2)$ algebra.\cite{remark}

 \section{Conclusion}

 To summarize, we have found that the energy level of the linearized vortex Dirac Hamiltonian in two dimensions, $E_N ^\pm = \pm \sqrt{2N}$, is $2N$-degenerate for $N \geq 1$. This degeneracy may be viewed as a consequence of an accidental $SU(2)$ symmetry and a supersymetry, which together lead to {\it two} multiplets, with the quantum numbers $j=N/2$ and $j=(N/2) -1$, being degenerate in energy. We have also obtained the spectrum and the degeneracies of the Hamiltonian in general dimension, showing in passing that there is always a non-degenerate zero mode.

 The bound-state spectrum obtained here may be observable by scanning tunneling microscopy, with the superconductivity in graphene or on a surface of a topological insulator induced by a proximity effect, for example, as a finite local density of states well below the bulk gap.

\section{Acknowledgement}

This work was supported by the NSERC of Canada (I. F. H.) and NSC Taiwan (C. K. L). I.F.H. is also grateful to the Max Planck Institute for the Physics of Complex Systems in Dresden for hospitality, and to Babak Seradjeh for useful discussions.


\begin{thebibliography}{99}
\bibitem{remark} It is easy to see that the only allowed trasformations under time reversal of three operators that close an $SU(2)$ algebra are a) all three odd, and b) one odd and the other two even.
\end{thebibliography}
\end{document}